\documentstyle[12pt,fleqn,epsfig]{article}
\begin{document}

\begin{center}

{\Large \bf Transverse Activity of Kaons  \\
 and the Deconfinement Phase Transition \\
 in  Nucleus--Nucleus Collisions \\ 
}

\vspace{1.5cm}

{\bf M.I. Gorenstein}$^{a,b}$,
{\bf M. Ga\'zdzicki}$^{c,d}$
and
{\bf K.A. Bugaev}$^{a,e}$,

\vspace{1cm} 

\noindent
\begin{minipage}[t]{12.5cm}  
$^a$ Bogolyubov Institute for Theoretical Physics,
Kiev, Ukraine\\
$^b$ Institut f\"ur Theoretische Physik, Universit\"at Frankfurt, Germany \\
$^c$ Institut f\"ur  Kernphysik, Universit\"at  Frankfurt,
Germany\\
$^d$ Swietokrzyska Academy, Kielce, Poland \\
$^e$ Gesellschaft f\"ur Schwerionenforschung (GSI), Darmstadt, Germany\\
\end{minipage} 

\end{center}

\begin{abstract}
\noindent
We found that the experimental results on transverse mass
spectra of kaons produced in central Pb+Pb (Au+Au) interactions
show an anomalous dependence on the collision energy.
The inverse slopes of the spectra increase with  energy
in the low (AGS) and high (RHIC) energy domains, whereas  they are constant
in the intermediate (SPS) energy range.
We argue that this  anomaly
is probably caused by a modification of the equation of state
in the transition region between confined and deconfined matter.
This observation may be considered  as a new signal, in addition to the
previously reported
anomalies in  the pion and strangeness production,
of the onset of deconfinement  located in the low SPS energy domain.
\end{abstract}

\newpage

The statistical model of the early stage, SMES, of nucleus-nucleus (A+A)
collisions suggests \cite{GaGo0} that the onset of
the deconfinement phase transition at the early stage of the collisions
may be signaled by the anomalous  energy dependence of several
hadronic observables.
In particular,
following earlier suggestions \cite{GaRo},
the behavior of strangeness and pion
yields in the transition region was studied in detail.
Recent
measurements \cite{na49} of pion and kaon
production in central Pb+Pb collisions at the CERN SPS indeed
indicate that the transient state
of deconfined matter is created in these collisions for
energies larger than about 40 A$\cdot$GeV.
The present data show a  maximum of the
strangeness to pion ratio at this  energy. An exact position and the
detailed structure of this maximum will be clarified
by the soon expected  new results from
the 2002 Pb run at 20 and 30~A$\cdot$GeV.

\vspace{0.2cm} In the present letter we discuss another well known
observable, which may be sensitive to the onset of deconfinement,
the transverse momentum, $p_T$, spectra of produced hadrons.
It was suggested by Van Hove
\cite{van-hove} more than 20 years ago to identify the deconfinement phase
transition in high energy proton--antiproton interactions by
an anomalous
behavior (a plateau-like structure) of the average transverse momentum as a
function of hadron multiplicity.
Let us briefly recall  Van Hove's arguments.
According
to the general concepts of the hydrodynamical approach the hadron
multiplicity reflects the entropy, whereas
the transverse hadron activity reflects
the combined effects of temperature and collective transverse expansion.
The entropy is assumed to be created at the early stage of the collision
and is approximately constant during the hydrodynamic expansion. The
multiplicity  is proportional
to the entropy, $S=s \cdot V$, where $s$ is the entropy density and
$V$ is the effective volume occupied by particles.
During the hydrodynamic expansion, $s$
decreases and $V$ increases with $s \cdot V$, being
approximately constant. The large multiplicity  at high energies means a
large entropy density at the beginning of the expansion (and consequently
a larger volume at the end).
A large
value of $s$ at the early stage of the collisions
means  normally high  temperature $T_{0}$ at this stage.
This, in
turn, leads to an increase of transverse hadron activity, a
flattening of the transverse momentum spectra.
Therefore, with
increasing  collision energy
\footnote{ In the original Van Hove
suggestion the correlation between average transverse momentum and hadron 
multiplicity was discussed for proton-antiproton collisions at fixed 
energy. Today we have the possibility to study  A+A collisions at different 
energies.}
 one expects to observe an increase of both the
hadron multiplicity and average transverse momentum per hadron. However, the
presence of the deconfinement phase transition would change this
correlation. In the phase transition region the initial entropy density
(and hence the final hadron multiplicity) increases with collision energy, 
but temperature $T_{0} = T_C$ and pressure  $p_{0} = p_C$
remain constant. 
The equation of state presented in the form
$p(\varepsilon)/\varepsilon$ versus 
energy density $\varepsilon$ shows a minimum (the 
`softest
point' \cite{SZ79,HS94}) at the boundary of the 
({\it generalized} \cite{HS94}) mixed phase and the quark gluon plasma. 
Consequently
the shape of 
the $p_T$ spectrum 
is approximately independent of the multiplicity or collision
energy.
The transverse 
expansion effect may even decrease when  crossing 
the transition region \cite{van-hove}. 
Thus one expects an anomaly in  
the energy dependence of transverse
hadron activity: the average transverse momentum increases with collision 
energy when the early stage matter is either in pure confined or in pure  
deconfined phases,
and it remains approximately constant when  the matter
is in the mixed phase.  


A simplified picture with $T=T_{C}=const$ inside the mixed phase is
changed if the created early stage matter has
a non--zero baryonic density.
It was however demonstrated \cite{HS97} that the main
qualitative features ($T\cong const$, $p\cong const$, 
and a minimum of the function
$p(\varepsilon)/\varepsilon$ $vs$ $\varepsilon$)
are present also in this case.
In the SMES model \cite{GaGo0}, which  correctly predicted the energy dependence
of pion and strangeness yields, the modification of the equation of
state due to the deconfinement phase transition
is located between 30 and  about 200 $A\cdot$GeV.
Thus the anomaly in energy dependence of transverse hadron activity
may be expected in this energy range.
Do we see this anomaly in the experimental data?

\vspace{0.2cm}
The experimental data  on transverse mass
($m_T = \sqrt{m^2 + p_T^2}$, where $m$ is a particle mass)
spectra are usually parameterized by a
simple exponential dependence:
\begin{equation}\label{exp}
\frac{dN}{m_{T}~dm_{T}}~=~C\exp\left(-\frac{m_{T}}{T^{*}}\right)~,
\end{equation}
where the inverse slope parameter $T^{*}$ 
is sensitive to both the thermal and
collective motion in the transverse direction. In the parameterization
(\ref{exp}) the shape of the $m_T$ spectrum is fully determined by
a single parameter, the
inverse slope $T^*$.
In particular,
the average transverse mass, $\langle m_{T} \rangle$,
can be expressed as:
\begin{equation}\label{mt}
\langle m_{T}\rangle ~=~ T^{*}~+~m~+~\frac{(T^{*})^{2}}{m~+~T^{*}}~.
\end{equation}

The energy dependence of the inverse slope parameter fitted to
the $K^+$ and $K^-$ spectra for central Pb+Pb (Au+Au)
collisions is shown in Figs. 1 and 2.
The results obtained at AGS \cite{ags}, SPS \cite{na49} and
RHIC \cite{rhic} energies are compiled.
The striking features of the data can be summarized and interpreted as
follows.
\begin{itemize}
\item
The $T^{*}$ parameter increases strongly with collision energy up to the lowest
(40 A$\cdot$GeV) SPS energy point.
This is an energy region where the creation of confined matter at
the early stage of the collisions is expected.
Increasing collision energy leads to an increase of the
early stage  temperature and pressure.
Consequently  the  transverse activity of produced hadrons,
measured by the inverse slope parameter, increases with increasing energy.
\item
The $T^{*}$ parameter is approximately independent
of the collision
energy in the SPS energy range.
In this energy region the transition between confined and deconfined matter
is expected to be located.
The resulting modification of the  equation of state 
``suppresses'' the hydrodynamical transverse expansion and
leads to the observed plateau structure in 
the energy dependence of the $T^*$ parameter.
\item
At higher energies (RHIC data)  the $T^{*}$ again increases with collision
energy. The equation of state  at the early stage  becomes again stiff,
the  early stage temperature and pressure  increase with collision energy.
This results in increase of $T^{*}$ with  energy.
\end{itemize}

\vspace{0.2cm}
The anomalous energy dependence of the $m_T$ spectra is a characteristic
feature of the kaon data.
Why is this the case?
How do the $m_T$ spectra of other hadrons look like?
The answer is rather surprising: among the measured hadron species the kaons
are the best and unique particles for observing the effect of the
modification of the equation of state due to the onset
of deconfinement.
The arguments are 
 as follows.
\begin{itemize}
\item
The kaon $m_{T}$--spectra are only weakly affected by the hadron
re--scattering and resonance decays during the post--hydrodynamic hadron
cascade \cite{BD,Sh}.
\item
A simple one parameter exponential fit (\ref{exp}) is quite accurate up to $m_{T}-m
\cong 1$~GeV for $K^{+}$ and $K^{-}$ mesons in A+A collisions at all
energies. This means that the energy dependence of the average transverse
mass $\langle m_{T} \rangle$  and average transverse momentum
$\langle p_{T} \rangle$ for kaons is
qualitatively the same as that for the parameter $T^{*}$.
This simplifies the analysis of the experimental data.
\item
The high quality data on $m_T$ spectra of
$K^+$ and $K^-$ mesons in central Pb+Pb (Au+Au)
collisions
are available in the full range of
relevant energies.
\end{itemize}

The ``hydro QGP + hadron cascade'' approach \cite{BD,Sh} predicts a
strong modification of the $m_{T}$-spectra of protons and lambdas during
the hadron cascade stage in A+A collisions at both the SPS and RHIC. As
the hadron gas expands, the pions excite $\Delta$ and $\Sigma^*$
resonances and transform some part of their transverse energy in the
nucleon and hyperon sectors. Therefore, the hadron re--scattering and
resonance decays lead to significant increase (about 40$\%$ \cite{Sh}) of
the inverse slope parameters $T^*$ for (anti)nucleons and (anti)lambdas at the
expense of the pion transverse energy 
(also see discussion in Ref.~\cite{BGG:02}). 
These changes of the slopes $T^{*}$
are not directly related  to the  equation of state
of the matter at the early stage.
It is rather difficult to separate these hadron-cascade effects and in
any case, this separation will be strongly model dependent.
Note also
that a simple exponential fit (\ref{exp}) neither works  for
$\pi$-mesons ($T^{*}_{low-p_{T}}>T^{*}_{high-p_{T}})$  \cite{pl}
nor for protons
and lambdas ($T^{*}_{low-p_{T}} < T^{*}_{high-p_{T}}$). This means that the
average transverse masses, $\langle m_{T}\rangle$, and their energy
dependence are not connected to the behaviour of the slope parameters in
the simple way described by Eq.
(\ref{mt}): one should separately consider both
$T^{*}_{low-p_{T}}$ and $T^{*}_{high-p_{T}}$ slopes for these hadrons (see
Ref.~\cite{Sh} for details).

The transverse activity of $\Omega$ hyperons and $\phi$ mesons should, 
as in the case of kaons,
be sensitive to the matter equation of state
at the early stage of the collisions.
These particles seem to decouple just after the
hydrodynamic expansion stops, they do not participate in the hadron
cascade stage \cite{BD,Sh,BGG:02,BGG:01}.
Unfortunately the spectra of  $\Omega$ hyperons are measured only at
top SPS and RHIC energies \cite{omega}.
More data exist for $\phi$ meson production \cite{phi}.
However, 
the
large uncertainties in the experimental results
do not allow to draw a definite conclusion on the
possible anomaly in the energy dependence of $m_T$
spectra. 

\vspace{0.2cm} In conclusion,
we observe an anomalous energy
dependence of  transverse mass
spectra of $K^+$ and $K^-$ mesons
produced in central Pb+Pb (Au+Au) collisions.
The inverse slopes of the $m_T$--spectra increase with  energy
in the AGS and RHIC energy domains, whereas they remain constant
in the intermediate SPS energy range.
We argue that this  anomaly
is  caused by a modification of the equation of state
in the transition region between confined and deconfined matter.
This observation may be considered  as a new signal, in addition to the
previously reported
anomalies in energy dependence of the pion and strangeness production,
of the onset of deconfinement  located at the low SPS energies.

\vspace*{0.4cm}

\noindent {\bf Acknowledgments.} We thank W. Greiner  and
R. Renfordt for discussions and
comments.
M. G.  has been partially supported by 
Bundesministerium fur Bildung und Forschung (Germany) and
the 
Polish Committee of
Scientific Research under grant 2P03B04123.


\newpage

\begin{figure}[p]
\epsfig{file=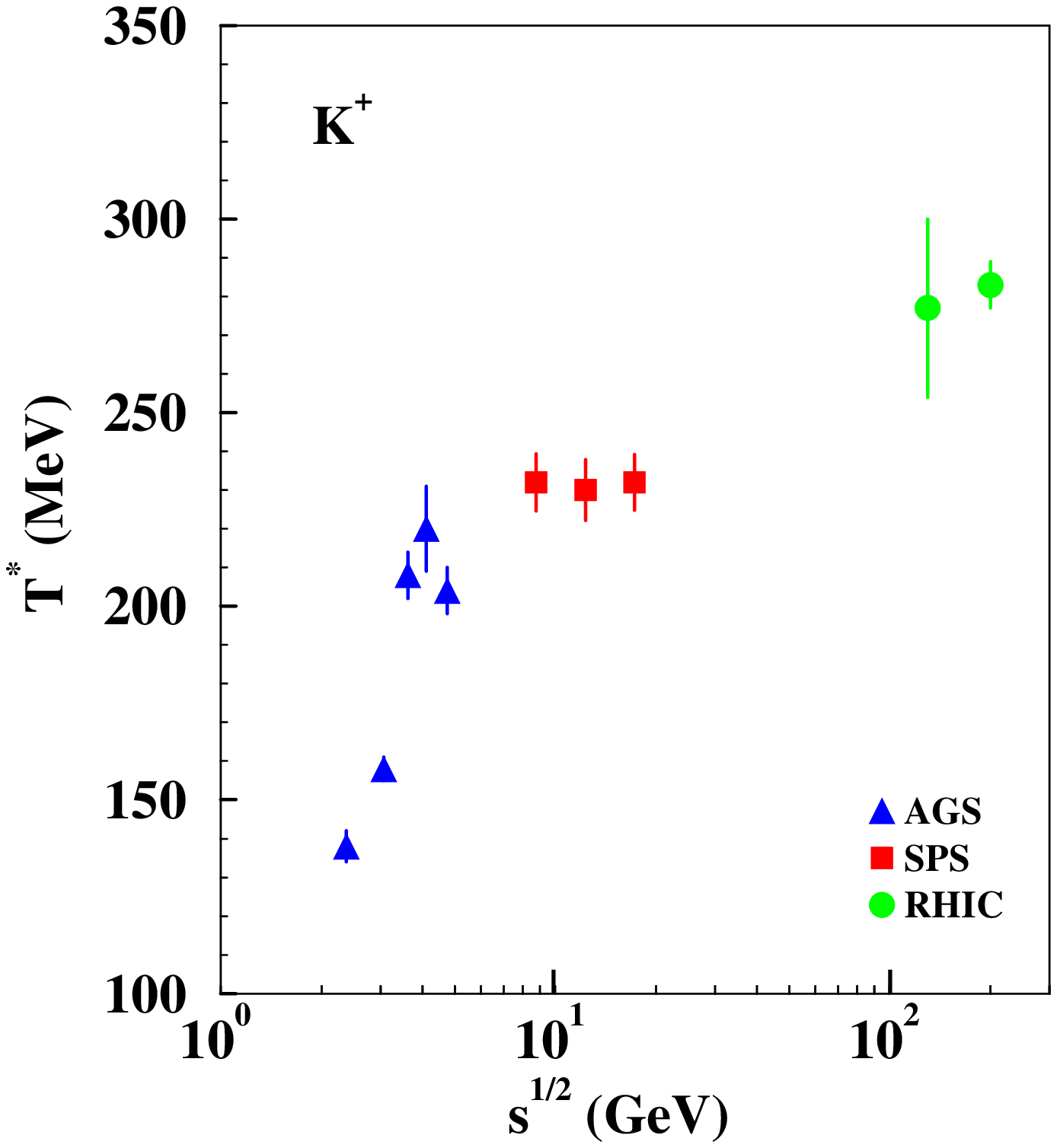,width=14cm}
\vspace{0.5cm}
\caption{
The energy dependence of the inverse slope parameter
$T^*$ for $K^+$ mesons produced at mid-rapidity in
central Pb+Pb (Au+Au) collisions at AGS 
\protect\cite{ags}
(triangles),
SPS 
\protect\cite{na49}
(squares) and RHIC 
\protect\cite{rhic}
(circles) energies.
}
\label{fig1}
\end{figure}

\newpage

\begin{figure}[p]
\epsfig{file=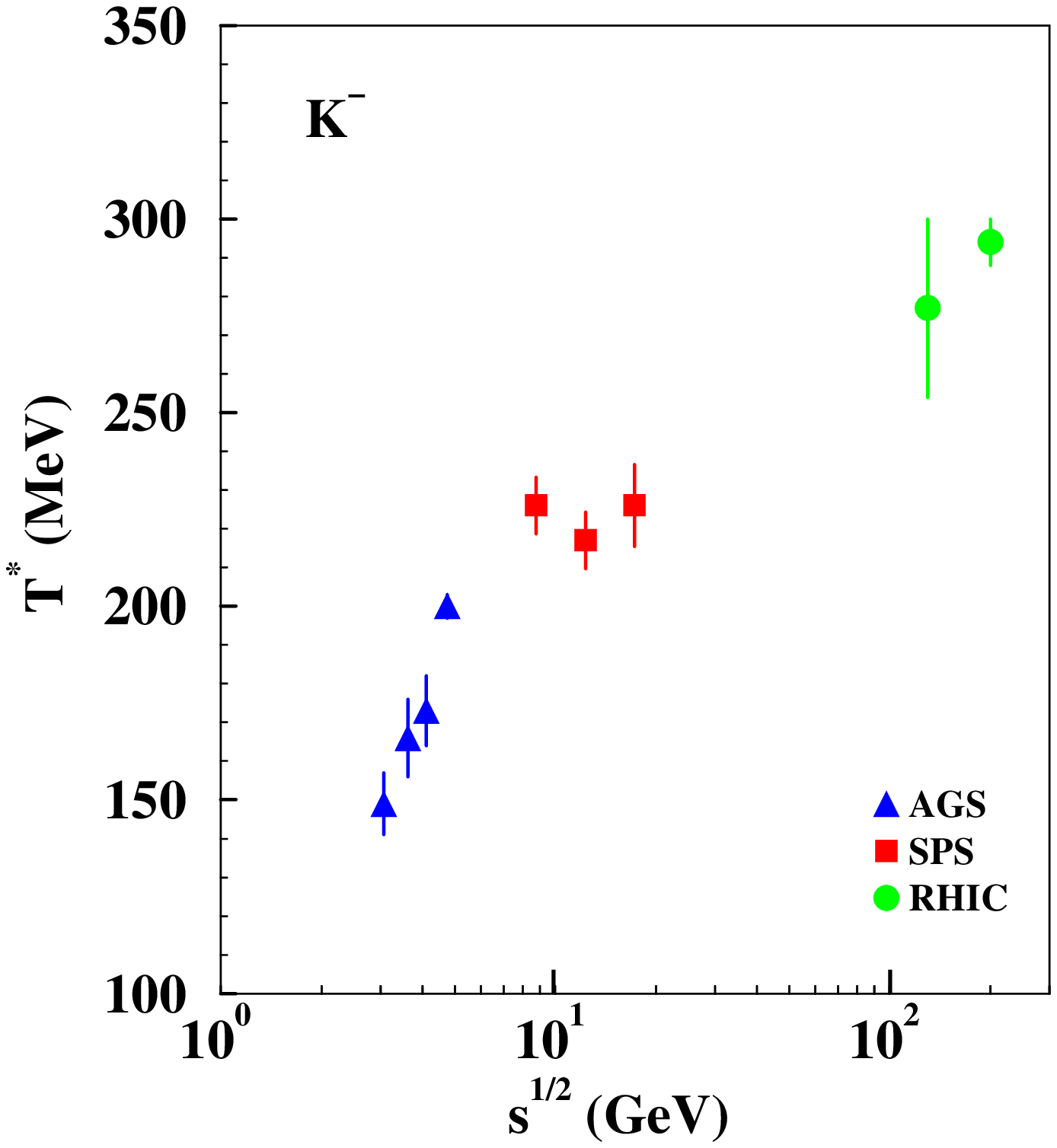,width=14cm}
\vspace{0.5cm}
\caption{
The energy dependence of the inverse slope parameter
$T^*$ for $K^-$ mesons produced at mid-rapidity in
central Pb+Pb (Au+Au) collisions at AGS 
\protect\cite{ags}
(triangles),
SPS 
\protect\cite{na49}
(squares) and RHIC 
\protect\cite{rhic}
(circles) energies.
}
\label{fig2}
\end{figure}

\end{document}